\documentclass[aps,pra,preprint,groupedaddress,longbibliography]{revtex4-1}
\usepackage{amsmath,amsfonts,amsthm}
\usepackage{graphicx}
\usepackage{xcolor}
\usepackage{physics}
\usepackage{hyperref}
\usepackage{bm}
\usepackage{ulem}
\begin{document}

\title{Density correlation functions and the spatial structure of the two-dimensional BEC-BCS crossover}

\author{J.C. Obeso-Jureidini and V. Romero-Roch\'in}
\affiliation{Instituto de F\'isica, Universidad Nacional Aut\'onoma de M\'exico \\
Apartado Postal 20-364, 01000 Cd. M\'exico, Mexico}

\date{\today}

\begin{abstract}

The spatial structure of a two-dimensional homogeneous mixture of fermionic atoms in two hyperfine states is analyzed throughout the BEC-BCS crossover. Within the BCS-Leggett mean-field model we consider three functions: the pair wave function and the density-density correlation functions between atoms of the same and of different hyperfine states. For the correlation functions we derive analytical expressions which allow to unveil the rich spatial structure of the superfluid. Mainly, we are able to study the large-distance behavior of the three functions, which exhibits an exponential decay and a well-defined oscillatory behavior. We report closed-form expressions for the correlation lengths and mean pair radius. Differences and similarities emerge when comparing with the three dimensional case. Particularly, we find an expression for the large-distance correlation length, in terms of the chemical potential and the gap,  valid in two and three dimensions, but whose dependence on the corresponding scattering lengths differ significantly.

\end{abstract}

\maketitle

\section{Introduction}
Experimental advances in ultracold atoms have surprised the community with the creation of physical systems with unprecedented control over their properties \cite{Hazra,Levinsen-Parish,Ketterle-review,Strinati-review,Bloch-review}.  An example is the achievement of spatial confinement to two dimensions (2D) \cite{Levinsen-Parish,Dyke_E,Sommer,Feld,Hueck-two-homogeneous,Lennart-observation-superfluidity}, offering an exceptional opportunity for testing theoretical models against experiments. Such is the case of the 2D BEC-BCS crossover,  implemented by Miyake \cite{Miyake}, and Randeria, Duan, and Shieh \cite{Randeria}. Their mean-field model, which concerns us, consists of a homogeneous balanced mixture of two fermionic species with a renormalized contact interaction between unlike particles, similarly to the three dimensional model of Leggett \cite{Leggett} and Eagles \cite{Eagles}. Depending on the strength of the interactions, different many-body states can be obtained. For weakly attractive interactions a gas of Cooper pairs is obtained, corresponding to the Bardeen-Cooper-Schrieffer regime (BCS) \cite{BCS-superconductivity}. On the other hand, for strongly attractive interactions a gas of diatomic molecules is generated, which condenses at zero temperature, corresponding to a Bose-Einstein condensation state (BEC). However, for intermediate strengths, the many-body states form a continuum that connects both regimes, known as the crossover region. An important question about the BEC-BCS crossover concerns the description of the pairing mechanism of unlike particles, giving rise to BCS-superfluidity and Bose-Einstein condensation \cite{Levinsen-Parish,Strinati-review}. To address this question we analyze the spatial structure of the gas within the 2D mean-field BCS-Leggett model aforementioned \cite{Miyake,Randeria}. The spatial structure can be extracted from the analysis of three functions: the correlation function between same species, the correlation function between different species, and the variational pair wave function.\\

The main contribution of this article is to present analytical expressions for the two correlation functions, an important result that allows to obtain a detailed picture of the behavior of the gas throughout the crossover. For the pair wave function, while we are unable to find an analytical expression for all values of the interaction, we do find its large-distance behavior, and we analyze it with the aid of numerical calculations. Also, we report closed-form expressions for the correlation lengths and mean pair radius defined as the respective second moment for each distribution. Similarly to the 3D case \cite{Obeso-Romero,Ortiz}, we can analyze the pairing phenomenon throughout the crossover. In the BCS limit, correlated pairs of unlike particles can have multiple sizes, particularly of macroscopic order, while Pauli-blocking correlations prevent two like particles to be found near each other. In the BEC limit the sizes of correlated pairs of unlike particles tend to be small, while Pauli-blocking correlations become negligible. Further, with our analytical expressions, we are able to study the large distance behavior of the correlation functions and the pair wave function. We find that the three functions exhibit an exponential decay and a well-defined oscillatory behavior.  The exponential decay, or large-distance correlation length, along the whole crossover, is characterized by the so-called 2D scattering length, a quantity that quantifies the two-body physics used to renormalize the atomic interaction strength. The spatial oscillation frequency turns out to be constant throughout the crossover, being equal to the Fermi wave number. These two results contrast interestingly with the 3D behavior \cite{Obeso-Romero}: while the large-distance correlation length is analogous for both dimensions in the BEC regime, but not in the crossover and the BCS side, the spatial oscillations agree in the BCS regime but are quite different in the BEC one. Nevertheless, it is remarkable that in both cases, 2D and 3D, the large-distance correlation length depends on the same way in terms of the thermodynamic variables chemical potential and gap, although their relationship to the corresponding scattering length differs.  Additionally, the existence of a finite large-distance correlation length allows to enquire into the anomalous breaking of the expected scale invariance of a 2D Fermi gas interacting through a contact potential \cite{Holten-etal-anomalous,Holstein_understanding,Holstein-pedestrians,Olshanii-etal_example_anomaly,Taylor-Randeria}. Indeed, such a correlation length determines a size of the density fluctuations, evidently being more notorious in the BEC regime and loosing its scale in the BCS side, as the fluctuations size grows without bound.\\

The article is organized as follows. In section \ref{sec_scattering} we give a brief review of the two-body scattering problem used to renormalize the interaction strength in the many-body problem. In section \ref{sec_mean-field}, we introduce the 2D mean-field BCS-Leggett model and discuss the relevance of the two-body physics. In section \ref{sec_pair-correlations} we present the main study of the two-body distributions. Analytical expressions for both correlation functions are reported. We analyze the large-distance behavior of the three functions by studying the spatial oscillation and the large-distance correlation length. Lastly, we report closed-form expressions for the correlation lengths and the mean pair radius. Final remarks are given in section \ref{sec_conclusions}.

\section{Two-body scattering properties for the many-body problem}\label{sec_scattering}

\begin{center}
\begin{figure}[h!]
\includegraphics[width=0.5\linewidth]{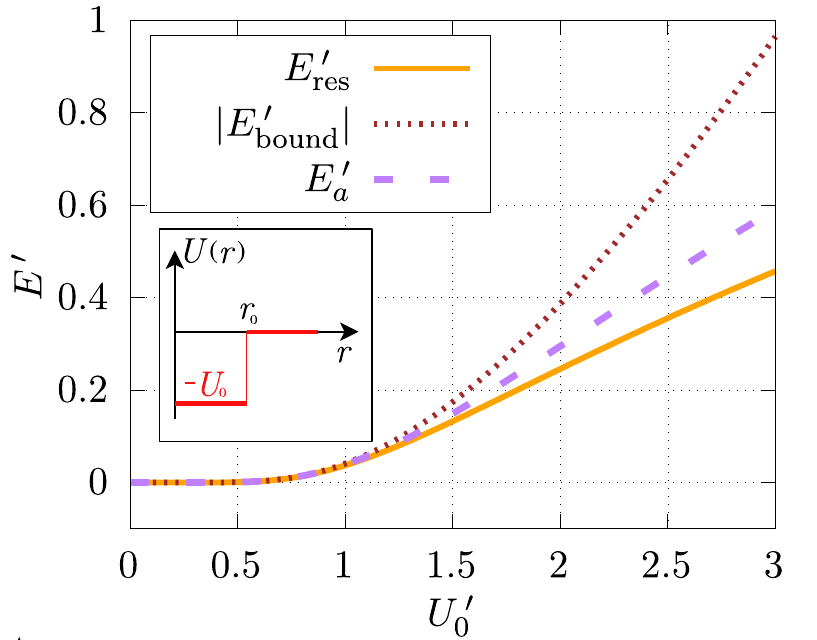}
\caption{(Color online) Comparison between the bound state energy $|E_\text{bound}|$, shown with a dotted line (brown); the exact s-wave resonance energy $E_{\text{res}}$ \cite{Adhikari}, solid curve (orange); and the approximate energy $E_a$, shown with a dashed line (purple), for the circular potential illustrated in the inset. The depth of the potential is given by $-U_0$ and the radius is $r_0$. The primed variables correspond to dimensionless quantities, using $m=\hbar=r_0=1$. } \label{fig_scattering_energies}
\end{figure}
\end{center}

In ultracold gases the complete coupled-channels interaction between atoms can be approximated by a single channel potential, facilitating the description of the physical properties \cite{Ketterle-review,Levinsen-Parish,Strinati-review,Kokkelmans-etal-res}. In this section we briefly revise the problem of scattering of equal mass particles from a non-divergent short-range central potential $U(r)$ with an attractive tail, and then, restrict ourselves to the ill-defined contact potential whose physical properties require a renormalization procedure \cite{Mead-Godines,Nyeo}.  Very generally, the low energy limit of the $T$ matrix is given by \cite{Randeria,Landau_qm,Adhikari,Lapidus}
\begin{equation}\label{eq_t_matrix-s}
T(2E) \approx \frac{4 \hbar^2}{m} \Bigg[  \frac{1}{ \text{ln}(E_a/2E)/\pi + i } \Bigg],
\end{equation}
where $m$ is the mass of a particle and $E= \hbar^2 k^2 /2m$ is half the energy of relative motion \cite{Randeria,Landau_qm}. The quantity $E_a$ is the approximate energy at which a scattering resonance occurs \cite{Randeria,Landau_qm,Adhikari,Lapidus}. However, when the potential $U(r)$ does not change sign in the interval $[0,r_0]$, where $r_0$ is the range of the potential, the energy $E_a$ is also a good approximation of the absolute value of an ever-present bound state energy $E_{\text{bound}}$ \cite{Landau_qm}, given by
\begin{equation}
E_a \approx \frac{4 \hbar^2}{m r_0^2} \text{exp}\Bigg(\frac{2}{\alpha_0} - 2 \gamma \Bigg),
\end{equation}
where $\gamma \approx 0.577$ is the Euler-Mascheroni constant and
\begin{equation}
\alpha_0 \approx \frac{m}{\hbar^2} \int_{0}^{r_0} dr \; r U(r).
\end{equation}
In Fig. \ref{fig_scattering_energies} we show a comparison between the energy of the $s$-wave resonance $E_{\text{res}}$ \cite{Adhikari, Lapidus}, the absolute value of the bound state energy $|E_{\text{bound}}|$, and the energy $E_a$, for a shallow-enough circular potential (analogous of the spherical well in 3D) of depth $-U_0$ and radius $r_0$. We can observe that for values $2mr_0^2U_0/\hbar^2 \lesssim 1$ the three energies are essentially equal. This result already indicates that $E_a$ is the scale that rules low energy collision physics in this interval.\\

A useful and appropriate model for the interaction between atoms in a many-body fluid is the contact interaction \cite{Strinati-review,Levinsen-Parish,Bloch-review}, which corresponds to a Dirac-delta function,
\begin{equation}
U(\mathbf{r})= g \delta^{(2)}(\mathbf{r}),
\end{equation}
where $g<0$ is the interaction strength. This potential presents several non-physical properties related to the zero range of the delta function \cite{Mead-Godines}. Therefore its use requires to renounce to describe high energy properties with accuracy \cite{Mead-Godines}. In order to calculate physical quantities, within a renormalization procedure, the interaction strength $g$ has to be expressed in terms of a given physical quantity such as the approximate energy of the $s$-wave resonance $E_a$ \cite{Mead-Godines}, or more generally in terms of the phase shift. This can be achieved by iterating the $T$ matrix in the Lippmann-Schwinger equation, from which an expression for the interaction strength in terms of the low energy limit of the $T$ matrix \eqref{eq_t_matrix-s} and a divergent term is obtained \cite{Strinati-review,Randeria,Ketterle-review,Salasnich-Toigo-zero,Morgan-Lee-Burnett},
\begin{equation}\label{eq_t-g-reg}
\begin{split}
 \frac{1}{g(\Lambda)}  &=  \frac{1}{T(2E)  } +  \frac{1}{A} \sum_{\mathbf{k}''}^{\Lambda}   \frac{1}{2(E - \varepsilon_{\mathbf{k}''} + i\delta)} \\
&= \frac{m}{4 \pi \hbar^2} \text{ln} \Bigg( \frac{E_a}{2 \Lambda}  \Bigg),
\end{split}
\end{equation}
where $A \rightarrow \infty$ is an auxiliary area and $\Lambda$ is a high energy cutoff. In the last equality we took the limit $\delta \rightarrow 0 ^+$ and neglected  terms smaller than the cutoff $\Lambda$. The divergence associated with the limit $\Lambda \rightarrow \infty$ allows to cancel out divergences that arise within the two-body problem \cite{Mead-Godines} and removes a divergence in the many-body problem, as shown below in section \ref{sec_mean-field}. Equation \eqref{eq_t-g-reg} can also be obtained from the bound state problem of the contact interaction, allowing to identificate $E_a = |E_{\text{bound}}|$, within this approximation \cite{Marini,Miyake,Salasnich-cond,Nyeo}. In the same fashion, the energy $E_a$ gives rise to a characteristic length of the interaction that can be identified as the 2D $s$-wave scattering length $a_{2\text{D}}$, though this definition differs from its analogue in 3D \cite{Werner}. In any case, for the sake of simplicity we will use the definition \cite{Levinsen-Parish,Mora}
\begin{equation}\label{eq_scattering_length}
a_{2\text{D}} = \Bigg(\frac{\hbar^2}{m E_a}\Bigg)^{1/2}.
\end{equation}
Note that for this 2D problem, this scattering length is always positive. As mentioned above, a bound state as well as a resonance are always present, but we can simply conclude that as the energy of the bound state vanishes the scattering length is very large and viceversa. The former leads to the BCS limit and the later to the BEC one.\\

Before proceeding to the many body problem, we point out an interesting and important aspect of the contact interaction model, namely, the anomalous breaking of scale invariance \cite{Holten-etal-anomalous,Holstein_understanding,Holstein-pedestrians,Olshanii-etal_example_anomaly,Taylor-Randeria}. This symmetry is expected for the contact potential since the Schr\"odinger equation of the relative motion is invariant under the transformation $\mathbf{r} \rightarrow \lambda \mathbf{r}$ and $E \rightarrow E / \lambda^2$ \cite{Holten-etal-anomalous}. However, this symmetry also implies the existence of an infinitely negative bound state energy, an unphysical situation \cite{Holstein_understanding}, and a constant $s$-wave phase shift \cite{Holstein-pedestrians}. Since during the renormalization procedure the problem must have a well-defined cross section, the approximate energy of $s$-wave resonance $E_a$ must also be introduced, making the physical properties no longer scale invariant \cite{Olshanii-etal_example_anomaly}. As we will discuss throughout the following sections, this same anomalous breaking occurs in the many body problem.

\section{Mean-field thermodynamics of an attractive 2D Fermi gas}\label{sec_mean-field}

\begin{center}
\begin{figure}[h!]
\includegraphics[width=0.5\linewidth]{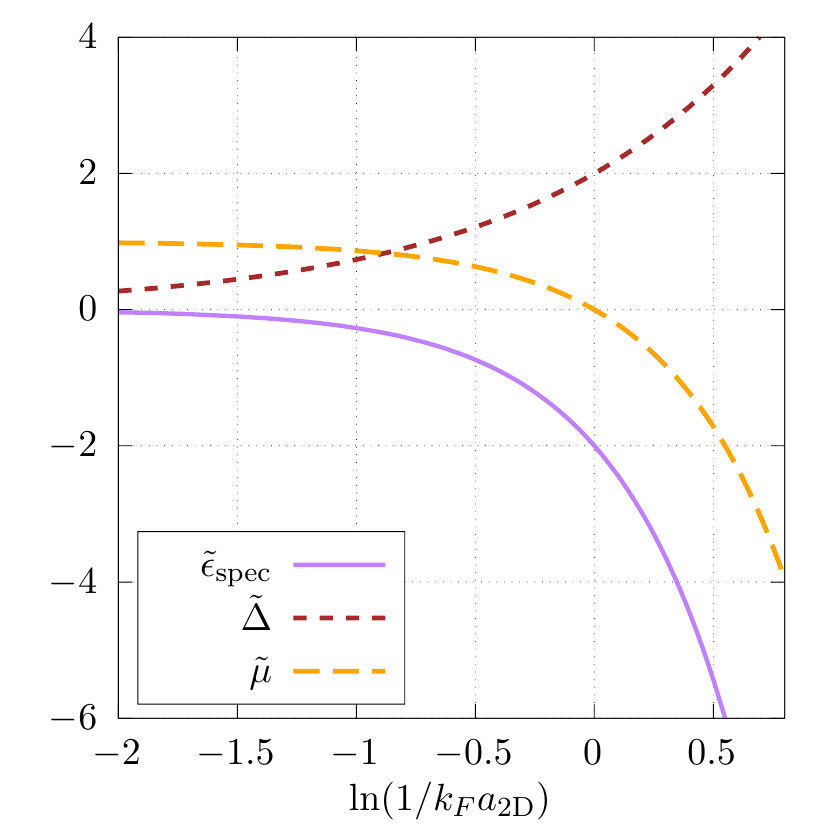}
\caption{ Dimensionless chemical potential $\tilde{\mu} = \mu / \epsilon_F$, gap $\tilde{\Delta} = \Delta / \epsilon_F$ and spectroscopic threshold energy $\tilde{\epsilon}_{\text{spec}}= \epsilon_{\text{spec}}/ \epsilon_F $. The BCS limit corresponds to $\text{ln}(1/k_F a_{2 \text{D}}) \rightarrow -\infty$, while the BEC limit is $\text{ln}(1/k_F a_{2 \text{D}}) \rightarrow \infty$. The spectroscopic energy obeys $\epsilon_{\text{spec}}= \epsilon_b = E_a$, with $\epsilon_b$ the binding energy per pair, see the text.} \label{fig_mu_delta}
\end{figure}
\end{center}

We consider a balanced gas mixture of fermionic atoms of mass $m$ in two hyperfine states, interacting via a contact potential. In the following we will treat these states as spins with the notation $\sigma = \uparrow, \, \downarrow$. In the low-density and low-energy limit we consider the grand potential $\hat{\Omega} = \hat{H}- \mu \hat{N}$ given by
\begin{equation}
\hat{\Omega} = \sum_{\mathbf{k},\sigma} (\epsilon_{k}-\mu)c_{\mathbf{k}\sigma}^{\dagger}c_{\mathbf{k}\sigma} + \frac{g}{A} \sum_{\mathbf{k}_1 \mathbf{k}_2} c_{\mathbf{k}_1 \uparrow}^{\dagger} c_{-\mathbf{k}_1 \downarrow}^{\dagger} c_{-\mathbf{k}_2 \downarrow} c_{\mathbf{k}_2 \uparrow},
\end{equation}
where $\epsilon_k = \hbar^2 k^2 /2m$, $g$ is the interaction strength and $A$ is the area of the sample. The operator $c_{\mathbf{k}\sigma}^{\dagger}$ creates a fermionic atom with momentum $\mathbf{k}$ and spin $\sigma$, the sums are over all wave vectors $\mathbf{k}$. The Hamiltonian $\hat{H}$ exhibits scale invariance due to the two-dimensional contact potential \cite{Holten-etal-anomalous}. However,  as mentioned above, this symmetry will be anomalously broken when expressing the interaction strength $g$ in terms of $E_a$ during renormalization, see equation \eqref{eq_t-g-reg}. As in 3D, the ground state energy can be estimated by means of the mean-field method \cite{Salasnich-cond,Tinkham,Randeria,Miyake} or with the BCS-Leggett variational method \cite{Leggett}, which introduces the BCS wave function \cite{BCS-superconductivity}
\begin{equation}\label{eq_bcs_state}
\ket{\Psi_{\text{BCS}}} = \prod_{\mathbf{k}} (u_{k} + v_{k} c_{\mathbf{k}\uparrow}^{\dagger} c_{-\mathbf{k}\downarrow}^{\dagger}) \ket{0},
\end{equation}
where the variational parameters satisfy the normalization condition $|u_k|^2 + |v_k|^2=1$. As usual, the number equation is given by
\begin{equation}\label{eq_number_equation}
n = \frac{1}{A} \sum_{\mathbf{k}} \Bigg( 1 - \frac{\epsilon_k - \mu }{\sqrt{(\epsilon_k - \mu )^2 + \Delta^2 }}  \Bigg),
\end{equation}
where $n= N/A$ is the particle density, which defines the Fermi wave number $k_F = \sqrt{2 \pi n}$. A thermodynamic quantity that arises naturally from the minimization procedure is the gap $\Delta$ \cite{Romero_contact}, which is related directly to the interaction between fermions of unlike species by means of the gap equation
\begin{equation}\label{eq_gap_equation}
1 = - \frac{g(\Lambda)}{2A } \sum_{\mathbf{k}}^{\Lambda} \frac{1}{\sqrt{(\epsilon_k - \mu )^2 + \Delta^2}},
\end{equation}
where we introduced a cutoff $\Lambda$ and the renormalized interaction strength $g(\Lambda)$, see equation \eqref{eq_t-g-reg} \cite{Miyake,Randeria,Salasnich-cond}. As pointed by Randeria, Duan and Shieh, it is necessary to identify $E_a$ with the two-body bound state energy $|E_{\text{bound}}|$ \cite{Randeria}. The variational parameters can be expressed in terms of the gap and the chemical potential:
\begin{equation}
\left\{ \begin{array}{c}
u_k^2 \\ v_k^2
\end{array}\right\} = \frac{1}{2} \left[1 \pm \frac{\epsilon_{k}-\mu}{\sqrt{(\epsilon_{k}-\mu)^2 + \Delta^2}} \right].\label{uvk}
\end{equation}
With equation \eqref{uvk} the ground state energy per area $E_0/A$ can be calculated directly \cite{Salasnich-gaussian,Salasnich-Toigo-zero}
\begin{equation}\label{eq_ground_energy}
\begin{split}
\frac{E_0}{A} &=  \frac{1}{A}  \sum_{\mathbf k} \Bigg(\epsilon_{k} - \frac{\epsilon_{k}(\epsilon_{k}-\mu) - \Delta^2/2}{\sqrt{(\epsilon_{k}- \mu)^2 + \Delta^2}} \Bigg) \\
&= \frac{m}{4 \pi \hbar^2} \Bigg(  3 \mu \sqrt{\mu^2 + \Delta^2} + \mu^2 - \frac{\Delta^2}{2} \Bigg).
\end{split}
\end{equation}
Note that this expression is obtained without renormalization, similarly to the 3D case \cite{Obeso-Romero,Ortiz}. However, to obtain $\mu$ and $\Delta$ in terms of the density $n$ and the characteristic energy $E_a$, one must solve the number and gap equations \eqref{eq_number_equation} and \eqref{eq_gap_equation}.  Following the renormalization procedure of Ref. \cite{Randeria}, for instance, of the gap and number equations, one finds a deceivingly very simple result for the chemical potential $\mu$ and the gap $\Delta$  in terms of the density $n$ and the energy $E_a$  \cite{Miyake},
\begin{equation}\label{eq_chem_ea}
\mu  = \epsilon_F - \frac{E_a}{2},
\end{equation}
and
\begin{equation}\label{eq_gap_ea}
\Delta = \sqrt{2 \epsilon_F E_a},
\end{equation}
where $\epsilon_F = \hbar^2 k_F^2 / 2m$ is the Fermi energy. In turn, the ground state energy, equation \eqref{eq_ground_energy}, can be readily expressed in terms of  $N$ and $E_a$ as
\begin{equation}\label{eq_ground_state_renormalized}
E_0 = E_F - N \frac{E_a}{2},
\end{equation}
where $E_F = N \epsilon_F /2$ is the energy of the non-interacting system, namely of an ideal gas of $N$ spin $1/2$ fermions. Recalling the definition of the scattering length $a_{2 \text{D}}$, equation \eqref{eq_scattering_length}, one finds that indeed, the BCS limit corresponds to $a_{2 \text{D}} \rightarrow \infty$ or $E_a \rightarrow 0$, and the BEC extreme to $a_{2 \text{D}} \rightarrow 0$ or $E_a \rightarrow \infty$ \cite{Miyake,Randeria}. This will be of relevance in our discussion of the correlation functions below. In principle, the thermodynamics of the gas at zero temperature may be obtained from $E_0$,  eq. \eqref{eq_ground_state_renormalized}. As argued by Werner and Castin \cite{Werner}, it is convenient to use the variable $\text{ln}(1/a_{2\text{D}})$ instead of $E_a$, such that the BCS limit corresponds to $\text{ln}(1/k_Fa_{2\text{D}}) \rightarrow - \infty$, while the BEC limit is $\text{ln}(1/k_Fa_{2\text{D}}) \rightarrow  \infty$. Additionally, its conjugate variable is related to  Tan's contact $C=m^2 \Delta^2 / \hbar^4$ \cite{Tan,Werner,Romero_contact,Vudtiwat},
\begin{equation}
\Bigg( \frac{\partial E_0}{\partial ( \text{ln} \; a_{2\text{D}} )} \Bigg)_{A,N} = \frac{\hbar^2 C A}{2 \pi m}.
\end{equation}
In  Fig. \ref{fig_mu_delta} we illustrate the behavior of $\mu$, $\Delta$ and $E_0$ across the BEC-BCS crossover, showing a qualitative agreement with the corresponding behavior in 3D. Another thermodynamic quantity 
relevant for the study of the anomalous breaking of scale invariance is the pressure $p$, given by
\begin{equation}
p = \frac{\pi \hbar^2}{2m} n^2.
\end{equation}
This is a striking result since it is also the pressure of a non-interacting Fermi gas. That is, the pressure of the interacting gas is independent of the interaction energy $E_a$. This result in turn indicates that the adiabatic compressibility has a scale-invariant behavior, as discussed in Ref. \cite{Taylor-Randeria}, thus showing an apparent agreement with the expected scale invariance. However, it is important to point here that corrections beyond mean-field remove the scale invariance in the adiabatic compressibility, as can be deduced from the thermodynamic results of Ref. \cite{Salasnich-gaussian}. An important observation is that in the BCS limit the small value of $E_a$ makes an apparent recovery of scale invariance, see equations \eqref{eq_chem_ea} and \eqref{eq_gap_ea}. In agreement with this apparency, the density-density correlation functions exhibit a scale-invariance behavior on the BCS side, as shown in the next section.

For further purposes below, let us introduce two energies that are related to the binding properties of pairs of opposite spin. The first one is the threshold energy required to create a quasiparticle by exciting an atom to a third state with negligible momentum transfer $ \epsilon_{\text{spec}} $ given by \cite{Ketterle-review,schunckdetermination,Sommer}
\begin{equation}\label{eq_epsilon_spec}
\epsilon_{\text{spec}} =  \sqrt{\mu^2 + \Delta^2} - \mu.
\end{equation}
In the next section it will be seen that this quantity determines the large-distance exponential decay, while its 3D analogue $\epsilon
_{\text{spec}}^{3\text{D}}$, which has the same expression of equation \eqref{eq_epsilon_spec} \cite{schunckdetermination,Obeso-Romero}, also determines the 3D exponential decay. Therefore, an important role will be given to $\epsilon_{\text{spec}}$, emphasizing the expression in equation \eqref{eq_epsilon_spec} in terms of $\mu$ and $\Delta$.  In Fig. \ref{fig_mu_delta} we also show $\epsilon_{\text{spec}}$ for  comparison with the two other relevant energies.  The other quantity to be introduced is the binding energy per pair $\epsilon_b$, given by \cite{Randeria,Sommer}
\begin{equation}
\epsilon_b = \frac{2}{N} (E_F - E_0) .
\end{equation}
By inspection of equations \eqref{eq_chem_ea}-\eqref{eq_ground_state_renormalized}, we find that both energies are equal to $E_a$, namely,
\begin{equation}\label{eq_equal_energies}
\epsilon_{\text{spec}} = \epsilon_b = E_a.
\end{equation}
This is in stark contrast with the 3D BEC-BCS system where it was found that $\epsilon_b^{3\text{D}} \neq \epsilon_{\text{spec}}^{3\text{D}}$ \cite{Obeso-Romero,Ortiz}.\\

\section{Pair wave function and density correlation functions}\label{sec_pair-correlations}

The two-body functions we analyze in this study are relevant for understanding the spatial structure of the gas and have been a subject of interest for ultracold gases \cite{Landau_statistical,Giorgini-Stringari,Ketterle-review,Strinati-review,Kadin,Chuanzhou,Murthy_bkt,Vudtiwat}. For instance, information about the probability of finding two types of particle at different spatial points can be extracted from the density-density correlation functions. Also, they measure the relation between density fluctuations at different points, and yield their characteristic length scales.  These are,
\begin{equation}
G_{\sigma \sigma '} (\mathbf{x}, \mathbf{x}^{\> \prime}) = \langle \hat n_\sigma(\mathbf{x}) \hat n_{\sigma '}(\mathbf x^{\> \prime}) \rangle -\langle \hat n_\sigma(\mathbf{x}) \rangle \langle  \hat n_{\sigma '}(\mathbf x^{\> \prime}) \rangle,
\end{equation}
where $\sigma$ and $\sigma '$ are spin labels which can take the values $\uparrow$ or $\downarrow$. We have introduced the particle density operator at point $\mathbf{x}$ of spin $\sigma$ given by $\hat{n}_{\sigma}(\mathbf{x}) = \hat{\psi}_{\sigma}^{\dagger}(\mathbf{x}) \hat{\psi}_{\sigma}(\mathbf{x})$, where
\begin{equation}
\hat{\psi}_{\sigma}(\mathbf{x}) = \frac{1}{\sqrt{A}} \sum_{\mathbf{k}} e^{i \mathbf{k} \cdot \mathbf{x}} \; c_{\mathbf{k} \sigma}
\end{equation}
are the usual field operators. The equal population property $N_{\uparrow} = N_{\downarrow}$ allows us to equal the labels as $\uparrow \uparrow = \downarrow \downarrow$ and $\uparrow \downarrow = \downarrow \uparrow$.  Within the BCS theory, these correlations are given by
\begin{equation}\label{eq_antiparallel-corr-int}
G_{\uparrow \downarrow} (\mathbf{r}) = |  g_{\uparrow \downarrow} (\mathbf{r}) |^2,
\end{equation}
and
\begin{equation}\label{eq_parallel-corr-int}
G_{\uparrow \uparrow} (\mathbf{r}) = \frac{n}{2} \delta^{(2)}(\mathbf{r}) - |  g_{\uparrow \uparrow}(\mathbf{r}) |^2
\end{equation}
where we have defined
\begin{equation}
g_{\uparrow \downarrow} (\mathbf{r}) = \frac{1}{(2 \pi)^2} \int d^2 k \; e^{i \mathbf{k} \cdot \mathbf{r}} \; u_k v_k \>,
\end{equation}
and
\begin{equation}
g_{\uparrow \uparrow} (\mathbf{r}) = \frac{1}{(2 \pi)^2} \int d^2 k \; e^{i \mathbf{k} \cdot \mathbf{r}} \; v_k^2 \>,
\end{equation}
with $\mathbf{r}=\mathbf{x}-\mathbf{x}^{\> \prime}$.  The latter two quantities are the one-body Green functions of this problem.\\
On the other hand, the spatial projection of the BCS state \eqref{eq_bcs_state} to a fixed number of particles state allows us to identify the unnormalized pair wave function \cite{Leggett}:
\begin{equation}\label{eq_pair_w_function}
\phi_{\text{BCS}} (\mathbf{r}) = \frac{1}{(2\pi)^2} \int d^2 k\; e^{i \mathbf{k} \cdot \mathbf{r}} \; \frac{v_k}{u_k}.
\end{equation}

\begin{center}
\begin{figure}[h!]
\includegraphics[width=0.8\linewidth]{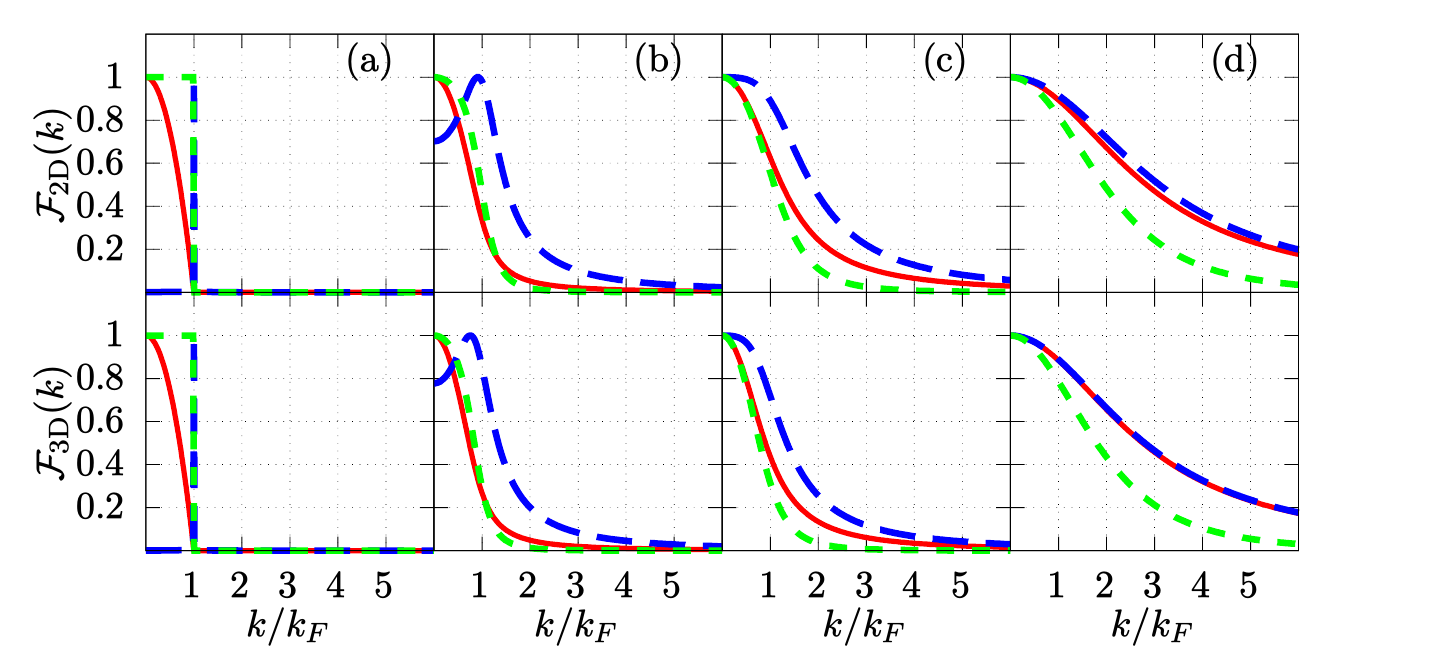}
\caption{(Color online) Behavior of $\mathcal{F}(k) = v_k/u_k$, solid (red) line; $\mathcal{F}(k)=v_k u_k$, long (blue) dashes; and $\mathcal{F}(k)=v_k^2$ short (green) dashes. The upper panel corresponds to the 2D system, while the lower panel to the 3D case. All curves are normalized such that their maximum is equal to one. Column (a) is at the BCS limit, where $\text{ln}(1/k_F a_{2\text{D}}) \approx -8.49$, $1/k_F a_{3\text{D}} \approx -5$. Column (b) is by the crossover, where $\text{ln}(1/k_F a_{2\text{D}}) \approx -0.891$, $1/k_F a_{3\text{D}} \approx 0.028$. Column (c) is where the chemical potential is zero, $\text{ln}(1/k_F a_{2\text{D}}) \approx 0.023$, $1/k_F a_{3\text{D}} \approx 0.578$. Column (d) is at the BEC limit, $\text{ln}(1/k_F a_{2\text{D}}) \approx 1.009$, $1/k_F a_{3\text{D}} \approx 2.728$.} \label{fig_vksuks}
\end{figure}
\end{center}

 As we can observe, the above physical quantities are given by the variational parameters $v_k$, $u_k$ \cite{Strinati-review,Ketterle-review}. For the purposes of analyzing the 2D and 3D differences of the spatial structure of the gas throughout the crossover, we find it interesting to compare the behavior of $v_k/u_k$, $v_k u_k$ and $v_k^2$. In Fig.  \ref{fig_vksuks}  the upper panel corresponds to 2D, while the lower panel to the 3D case, at characteristic thermodynamic states, namely, at the far BCS regime, very near the crossover $1/k_F a_{3D} = 0$ in 3D and $\ln(1/k_F a_{2D}) = 0$, at $\mu = 0$, and at the deep BEC side. In wave vector space it is seen that the behavior of the variational parameters is qualitatively similar between 2D and 3D \cite{Obeso-Romero,Giorgini-Stringari,Strinati-review}. However, as we will see below, their spatial behavior obtained by Fourier transforms are quite different.

 To proceed we need to calculate  two-dimensional Fourier transforms of the form,
\begin{equation}
f(\mathbf{r}) = \int d^2 k \; e^{i \mathbf{k} \cdot \mathbf{r}} \; \mathcal{F}(\mathbf{k}),
\end{equation}
where $\mathcal{F}(\mathbf{k})$ can be $v_k / u_k$, $u_k v_k$ or $v_k^2$. For the 3D system the angular integrals can be simply evaluated and we can complete a one-dimensional Fourier transform. Instead, in 2D, performing the angular integral leads us to a Hankel transform:
\begin{equation}\label{eq_hankel_trans}
f(r) = 2\pi \int_0^{\infty} dk \, k \mathcal{F}(k) \, J_0(kr),
\end{equation}
where we used $f(\mathbf{r}) = f(r)$, $r= |\mathbf{r}|$, and we identified the integral representation of the Bessel function of the first kind of order zero \cite{Bateman}:
\begin{equation}\label{eq_besselj0}
J_0 (kr) = \frac{1}{2 \pi } \int_0^{2\pi} d\theta \; e^{ikr\cos \theta}.
\end{equation}
For the correlation functions, their respective integrals in equation \eqref{eq_hankel_trans} can be calculated analytically. The derivation for $g_{\uparrow \downarrow}(r)$ is relatively straightforward, but quiet lengthy for $g_{\uparrow \uparrow}(r)$. The details of this procedure are given in Appendix \ref{ap_inv_fourier_transf}. The final expressions of the correlation functions are:
\begin{equation}\label{eq_gud}
G_{\uparrow \downarrow}(r) = \left|\frac{m \Delta}{2 \pi \hbar^2} J_0(k_F r) \, K_0 \left(\frac{r}{\chi_{\text{spec}}}  \right)\right|^2,
\end{equation}
and
\begin{equation}\label{eq_guu}
G_{\uparrow \uparrow} (r) = \frac{n}{2} \delta^{(2)}(\mathbf{r}) - \left|\frac{m \Delta}{2 \pi \hbar^2} J_1(k_F r) \, K_1 \left(\frac{r}{ \chi_{\text{spec}} }  \right)\right|^2,
\end{equation}
where,
\begin{equation}\label{eq_chi_spec_2d}
\chi_{\text{spec}} = \Bigg( \frac{\hbar^2}{m \epsilon_{\text{spec}}} \Bigg)^{1/2} \>,
\end{equation}
which, as justified below, we define it as the large-distance correlation length. It is very important to recall that this length can be expressed in terms of $\mu$ and $\Delta$ by means of the threshold energy $\epsilon_{\text{spec}} = \sqrt{\mu^2+\Delta^2} - \mu$ in eq. \eqref{eq_epsilon_spec}. In 2D the large-distance correlation length is equal to the scattering length $\chi_{\text{spec}}= a_{2\text{D}}$, but we emphasize its definition in terms of $\epsilon_{\text{spec}}$ because this thermodynamic expression is also valid in 3D. However, to emphasize the anomalous breaking of scale invariance we will also use $a_{2\text{D}}$ instead of $\chi_{\text{spec}}$, in some limits. For the pair wave function $\phi_{\text{BCS}}(r)$ we were unable to find an analytical expression for all values of $r$. However, we will present a semi-analytical and numerical analysis of its large-distance behavior. Nevertheless, we can evaluate numerically equation \eqref{eq_hankel_trans} with $\mathcal{F}(k) = v_k /u_k$, a difficult task due to the Bessel function $J_0(kr)$ as it oscillates and decreases slowly as $kr \rightarrow \infty $.\\

\begin{center}
\begin{figure}[h!]
\includegraphics[width=0.8\linewidth]{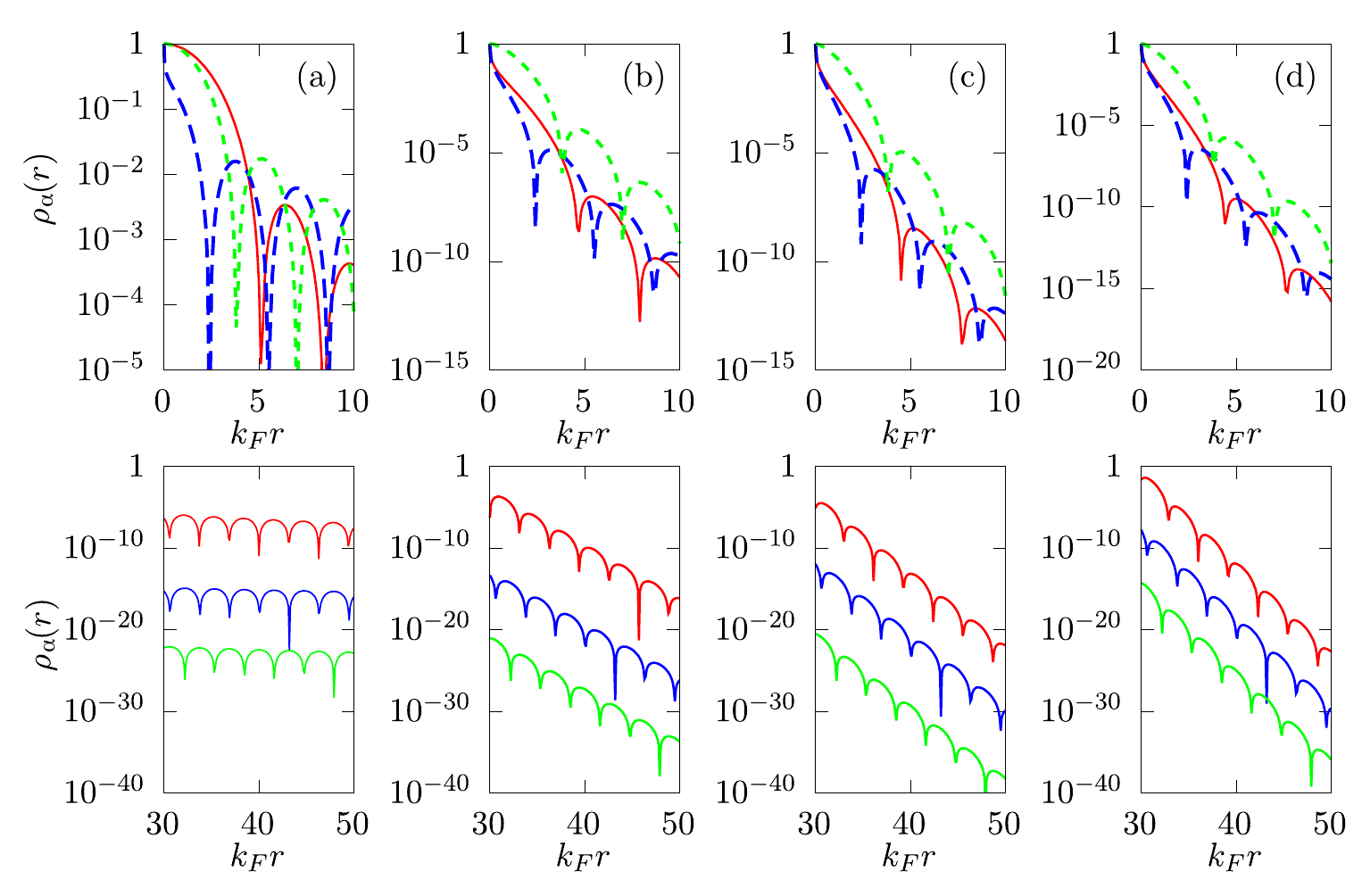}
\caption{(Color online) Correlation functions and pair wave function throughout the crossover. Each column corresponds to different points of the crossover, (a) $E_a /\epsilon_F = 0.0001$ (BCS side), (b) $E_a /\epsilon_F = 1$, (c) $E_a /\epsilon_F = 2$ (d) $E_a /\epsilon_F = 3$ (BEC side). In the upper panel the solid  line (red) corresponds to $\rho_\alpha(r) = |\phi_{\text{BCS}}(r)|^2$, the long dashed line (blue) to $\rho_\alpha(r) = G_{\uparrow \downarrow}(r)$ and the short dashed line (green) corresponds to $\rho_\alpha(r) = G_{\uparrow \uparrow}(r)$. They are plotted with arbitrary normalization. The lower panel shows the behavior at large distances, where the upper curves correspond to $|\phi_{\text{BCS}}(r)|^2$, the middle ones to $G_{\uparrow \downarrow}(r)$ and the lower curves to $-G_{\uparrow \uparrow} (r)$. In all the panels we removed the delta function of $G_{\uparrow \uparrow}(r)$. } \label{fig_panel}
\end{figure}
\end{center}

In Fig. \ref{fig_panel} the correlation functions and the pair wave function are plotted at characteristic thermodynamic states throughout the crossover. The upper panels show their short distance behavior while the lower ones the long distance spatial structure. Again, panel (a) is at the far BCS regime, (b) at the crossover, (c) at $\mu = 0$ and (d) at the deep BEC side.

In the BCS limit, $\text{ln}(1/k_F a_{2\text{D}}) \rightarrow - \infty$, Fig. \ref{fig_panel} upper panel (a), the three functions show an algebraic decay at small distances away from the origin. For very short distances, $k_F r \ll 1$, we can find asymptotic expressions,
\begin{equation}\label{eq_gud_short}
G_{\uparrow \downarrow}(r) \approx  \Bigg[  \frac{m \Delta}{2 \pi \hbar^2} \text{ln}\Bigg(\frac{r}{a_{2 \text{D}} }\Bigg) \Bigg]^2,
\end{equation}
\begin{equation}\label{eq_guu_short}
G_{\uparrow \uparrow}(r) \approx  \frac{n}{2} \delta^{(2)}(\mathbf{r})- \Bigg(\frac{m \Delta}{4 \pi \hbar^2} \Bigg)^2 \Bigg[ k_F \, a_{2 \text{D}} + \frac{(k_F r)^2}{2}   \Bigg( \frac{1}{ k_F a_{2 \text{D}}} \text{ln}\Bigg(\frac{e^{\gamma -1/2} \; r}{2 a_{2 \text{D}}}\Bigg)  - \frac{k_F a_{2 \text{D}} }{4}  \Bigg) \Bigg]^2.
\end{equation}
Equation \eqref{eq_gud_short} shows the expected logarithmic divergence multiplied by Tan's Contact $C = m^2 \Delta^2 / \hbar^4$ \cite{Werner,Levinsen-Parish}. This can be contrasted with 3D, where the divergence at the origin is algebraic \cite{Werner}. Equation \eqref{eq_guu_short} shows the behavior of Pauli-blocking at short distances, the probability of finding a particle of the same spin diminishing as $[r^2 \text{ln}(r)]^2$. On the other hand, in 3D the probability diminishes slower as $r^2$ \cite{Werner}. In contrast from the general approach of Werner and Castin \cite{Werner}, where the regular part of the pair wave functions are demanded to behave like a zero-energy scattering state, equations \eqref{eq_gud_short} and \eqref{eq_guu_short} were obtained demanding a well behavior of the gap and chemical potential, which is achieved by means of a renormalized interaction strengh, see equation \eqref{eq_gap_equation}. The difference of the BCS-Leggett approach is explicitly seen in the BCS limit where the variational pair wave function $\phi_{\text{BCS}}(r)$ exhibits a different behavior as that postulated in Ref. \cite{Werner}, that is,
\begin{equation}\label{eq_asymp_bcs_wave-function}
\phi_{\text{BCS}}(r) \approx \frac{2 \varepsilon_F }{\pi \Delta} k_F^2 \frac{J_{2}(k_F r)}{(k_F r)^2}.
\end{equation}
This asymptotic form remains finite at the origin $r=0$, while its large-distance algebraic decay and oscillatory behaviors are independent of  $a_{2 \text{D}}$, in agreement with the apparent recovery of scale invariance in the BCS limit, as discussed in section \ref{sec_mean-field}.

Quite generally, in the far BCS limit, short and large distances, the equal pair correlation function can be approximated as
\begin{equation}\label{eq_f_corr_parallel_free}
G_{\uparrow \uparrow}(r) \approx \frac{n}{2} \delta^{(2)}(\mathbf{r}) -\Bigg[ n \frac{J_1(k_F r)}{k_F r}  \Bigg]^2,
\end{equation}
which corresponds to the correlation function of a non-interacting ideal gas of $N$ fermions, becoming scale invariant in this limit. This behavior is analogous to the 3D case \cite{Giorgini-Stringari}. Within this same limit $G_{\uparrow \downarrow}(r) \approx 0$, again as in an ideal gas.\\

On the BEC side, $\text{ln}(1/k_F a_{2\text{D}}) \rightarrow \infty$, the correlation functions and the pair wave function get localized, see Fig. \ref{fig_panel} (b), (c) and (d), indicating the formation of bosonic molecules and the gradual lost of Pauli-blocking correlations \cite{Obeso-Romero}. Comparing the behavior of $v_{k}u_{k}$ and $v_k/u_k$ on the deep BEC side, $ -\mu \gg \Delta$, see Fig. \ref{fig_vksuks} (d), it can be concluded that,
\begin{equation}
\phi_{\text{BCS}}(r) \approx g_{\uparrow \downarrow}(r),
\end{equation}
similarly to the 3D system \cite{Ortiz,Obeso-Romero}.\\

Observation of the lower panels of Fig. \ref{fig_panel} suggests a common characterization of the large-distance behavior of the distributions $|\phi_{\text{BCS}}(r)|^2$, $G_{\uparrow \downarrow}(r)$ and $G_{\uparrow \uparrow}(r)$. That is, for large distances $k_F r \gg 1$ these distribution functions can be written in the form
\begin{equation}\label{eq_distributions}
\rho_{\alpha} (r) \approx \frac{\text{const}}{r^2} \,  \exp \left(- \frac{ r}{\chi_\alpha} \right) \, \mathcal{P}_{\alpha}(\kappa_\alpha r + \varphi_{\alpha}),
\end{equation}
where $\alpha= \text{BCS}, \, \uparrow \downarrow , \uparrow \uparrow$. In this expression we define the exponential decay length $\chi_\alpha$ as the large-distance correlation length (this definition has to be multiplied by $\sqrt{2}$ to agree with the exponential decay length of Ref. \cite{Obeso-Romero} in 3D). We have also introduced a periodic function $\mathcal{P}_{\alpha}(\kappa_\alpha r + \varphi_{\alpha}) $ with wave length $2 \pi / \kappa_\alpha$ and phase $\varphi_\alpha$. In the following we deal separately with these quantities.  We also discuss the second moments of the distributions, also known as correlation lengths in the literature. 

\subsection{Large-distance correlation lengths}

Using well-known properties of the Bessel functions \cite{Gradshteyn} for large values of their arguments, the correlation functions (\ref{eq_gud_short}) and (\ref{eq_guu_short}) can be written as
\begin{equation}\label{eq_gud_large}
G_{\uparrow \downarrow}(r) \approx \frac{\text{const} }{r^2} \, \exp \left(- \frac{ r}{a_{2 \text{D}}} \right) \cos^2 \left( k_F r - \frac{\pi}{4} \right),
\end{equation}
\begin{equation}\label{eq_guu_large}
G_{\uparrow \uparrow}(r) \approx - \frac{\text{const}}{r^2} \,  \exp \left(- \frac{ r}{a_{2 \text{D}}} \right) \cos^2 \left( k_F r - \frac{3 \pi}{4} \right) \>,
\end{equation}
where we have already written $\chi_{\text{spec}}=a_{2\text{D}}$ to highlight the role of the 2D scattering length. Comparing with the proposal given in eq. (\ref{eq_distributions}), we readily identify $\chi_{\uparrow \downarrow} =\chi_{\uparrow \uparrow} = a_{2\text{D}}$, a result anticipated before. On the one hand, this quantity defines the asymptotic behavior along the whole crossover in contrast to the 3D case, where the dependence on the scattering length occurs at the BEC side only. On the other hand, it shows an evident manifestation of  the anomalous breaking of scale invariance providing a size for the density fluctuations. In addition, this large-distance exponential decay of the correlation functions establishes a relation with the ever-present bound state throughout the crossover. \\

\begin{center}
\begin{figure}[h!]
\includegraphics[width=0.8\linewidth]{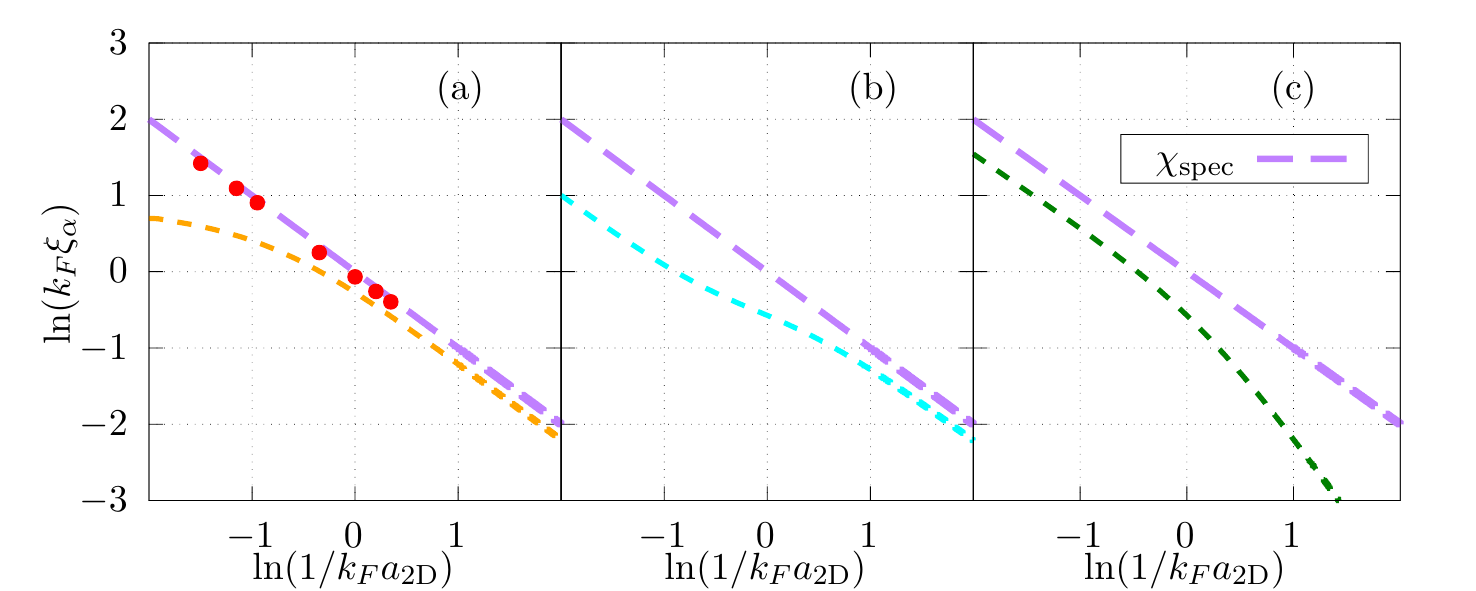}
\caption{(Color online) Characteristic lengths $\xi_{\alpha}$ of (a) the pair wave function $\phi_{\text{BCS}}(r)$, eq.  \eqref{eq_pair_w_function}, (b) the correlation function of opposite spins $G_{\uparrow \downarrow}(r)$, eq. \eqref{eq_antiparallel-corr-int} and (c) the correlation function of parallel spins $G_{\uparrow \uparrow}(r)$, eq.  \eqref{eq_parallel-corr-int}. In each panel, the short dashes correspond to (a) the mean pair radius $\xi_{\text{BCS}}$ and correlation lengths, (b) $\xi_{\uparrow \downarrow}$ and (c) $\xi_{\uparrow \uparrow}$, see eq. \eqref{eq_rms}. In all panels the large dashes (purple) correspond to the exponential decay length $\chi_{\text{spec}}=a_{2\text{D}}$, see eqs. \eqref{eq_equal_energies} and \eqref{eq_final_exp_length}. The dots (red) in (a) are numerical calculations of the exponential decay length $\chi_{\text{BCS}}$, showing good agreement with $\chi_{\text{spec}}=a_{2\text{D}}$. The correlation length $\xi_{\uparrow \downarrow}$ was reported in \cite{Randeria}.} \label{fig_lengths}
\end{figure}
\end{center}

For the pair wave function $|\phi_{\text{BCS}}(r)|^2$ we can extract its behavior by analyzing eq. \eqref{eq_hankel_trans} and from numerical calculations. To extract its large-distance behavior, we can approximate the Bessel function $J_0(kr)$, for $kr \gg 1$, by \cite{Marsden}
\begin{equation}\label{eq_aprox_Bessel}
J_0 (kr) \approx \frac{e^{i(kr- \pi /4) } + e^{-i(kr- \pi /4) }}{\sqrt{2 \pi k r}}.
\end{equation}
Then, similarly to 3D \cite{Obeso-Romero}, we can deform the corresponding Hankel transform integral in the complex plane to obtain the asymptotic behavior of the pair wave function. Details of this mathematical procedure are given in Appendix \ref{ap_large-distance-pair}. It is found that
\begin{equation}\label{eq_fr-theo}
\phi_{\text{BCS}}(r)  \propto \frac{e^{- r / a_{2\text{D}} }}{\sqrt{r}},
\end{equation}
thus identifying the large-distance correlation length $\chi_{\text{BCS}}= a_{2\text{D}}=\chi_{\text{spec}}$, as expected. To reinforce this conclusion we fitted the large-distance correlation length to the envelopes of $\phi_{\text{BCS}}(r)$, see lower panel in Fig. \ref{fig_panel}. These numerical values of $\chi_{\text{BCS}}$ correspond to the (red) dots in Fig. \ref{fig_lengths} (a), showing good agreement with $\chi_{\text{spec}}$, shown with long dashes in Fig. \ref{fig_lengths} (a).\\

It is thus concluded that the large-distance correlation lengths of equation \eqref{eq_distributions} are given by $\chi_{\text{spec}}$ or, equivalently, by the $s$-wave scattering length,
\begin{equation}\label{eq_final_exp_length}
\chi_{\alpha} = \chi_{\text{spec}} = a_{2\text{D}},
\end{equation}
for $\alpha= \text{BCS}, \, \uparrow \downarrow , \uparrow \uparrow$. Very differently from 3D, the scattering length $a_{2\text{D}}$ diverges in the BCS limit, consistently with the apparent recovery of scale invariance at the microscopic level, where fluctuation sizes increase, while in 3D the respective scattering length vanishes in the BCS limit, with the large-distance correlation length diverging.
For the purpose of associating the exponential decay behavior with a pair binding property we may define a \textit{pair-binding} function:
\begin{equation}\label{eq_binding-func}
|\Phi_{b}(r)|^2 = \frac{\text{const}}{r} \, e^{- 2 r / a_{2 \text{D}}}.
\end{equation}
This distribution has the large-distance asymptotic behavior of a bound state with zero angular momentum in a short-range central potential, as discussed in section \ref{sec_scattering}. Hence, the scattering length, being a two-body property associated to an ever-present bound state, determines the exponential decay behavior of the density fluctuations and of the pair wave function. We want to reemphasize that this two-body property is introduced by renormalization, contributing to the anomalous breaking of scale invariance.\\

We can now make a more precise comparison with the 3D system. In Ref. \cite{Obeso-Romero} it was shown that the 3D correlation functions $G_{\uparrow \downarrow}^{3\text{D}}(r)$, $G_{\uparrow \uparrow}^{3\text{D}}(r)$ and pair wave function $|\phi_{\text{BCS}}^{3\text{D}}(r)|^2$ show an exponential decay behavior at large distances of the form
\begin{equation}\label{eq_large_behavior3d}
\rho_{\alpha}^{3\text{D}}(r) \propto \frac{\text{const}}{r^2} \; \exp \left( -\frac{r}{\chi_{spec}^{3\text{D}}} \right) {\cal P}_{\alpha}^{3\text{D}}(r),
\end{equation}
where $\alpha=\text{BCS}, \, \uparrow \downarrow, \, \uparrow \uparrow$ and ${\cal P}_{\alpha}^{3\text{D}}(r)$ are periodic functions of $r$. In this expression we have introduced
\begin{equation}\label{eq_chi_3d}
\chi_{\text{spec}}^{3\text{D}} = \Bigg( \frac{\hbar^2}{m \, \epsilon_{\text{spec}}^{3\text{D}}} \Bigg)^{1/2},
\end{equation}
where $ \epsilon_{\text{spec}}^{3\text{D}} $ is defined as in equation \eqref{eq_epsilon_spec} but with the corresponding chemical potential and gap of a 3D system. Therefore, the threshold energy $\epsilon_{\text{spec}}$ strictly defined in terms of the gap $\Delta$ and the chemical potential $\mu$, as in equation \eqref{eq_epsilon_spec}, provides the large-distance correlation length $\chi_{\text{spec}}$ in 2D and in 3D, as shown by equations \eqref{eq_scattering_length}, \eqref{eq_equal_energies}, \eqref{eq_final_exp_length} and \eqref{eq_chi_3d}, although their values differ significantly. Independently, the large-distance behavior shown in equations \eqref{eq_large_behavior3d} and \eqref{eq_chi_3d} was reported in \cite{Pisani-Pieri-Strinati}.

\subsection{Characteristic wave numbers and phases}

\begin{center}
\begin{figure}[h!]
\includegraphics[width=0.7\linewidth]{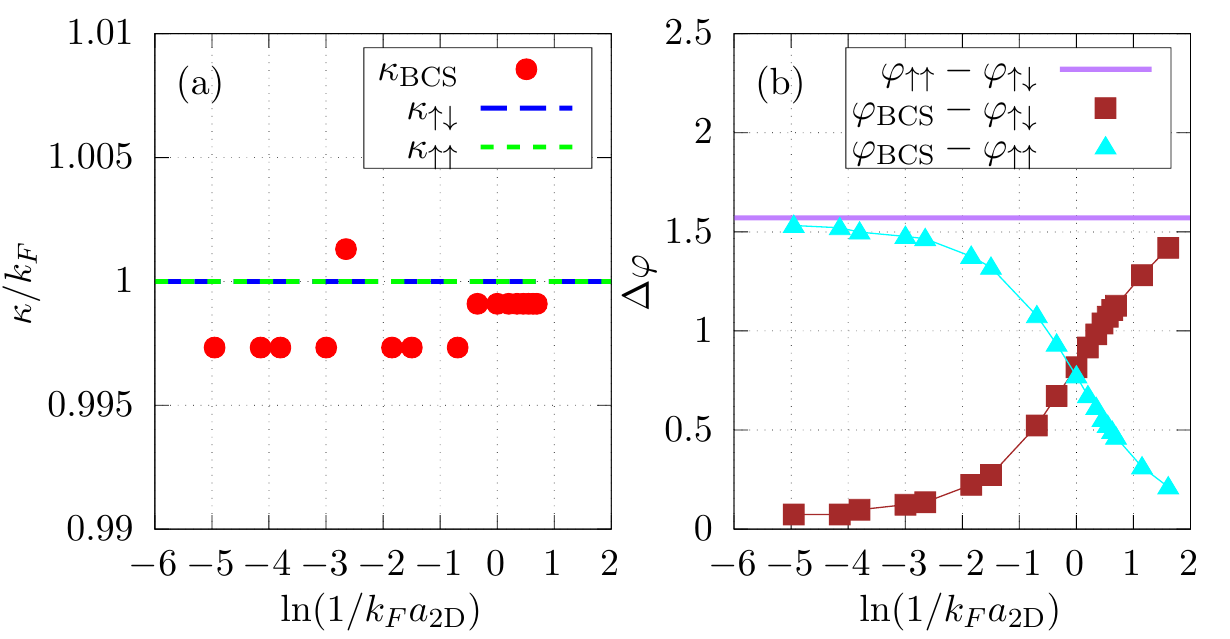}
\caption{(Color online) (a) Large-distance wave vectors $\kappa_{\text{BCS}}, \, \kappa_{\uparrow \downarrow}, \, \kappa_{\uparrow \uparrow}$ scaled with the Fermi wave number $k_F$, see eq. \eqref{eq_distributions}. The dots (red) correspond to numerical calculations of $\kappa_{\text{BCS}}$. The behavior $\kappa_{\uparrow \downarrow}= \kappa_{\uparrow \uparrow}= k_F$ was obtained analytically. (b) Phase differences between the pair wave function (BCS) and the density correlation functions ($\uparrow \uparrow, \, \uparrow \downarrow$). The phase difference $\varphi_{\uparrow \uparrow}-\varphi_{\uparrow \downarrow}$ is a theoretical result, while the other two were obtained numerically (the lines are guides to the eye).} \label{fig_phases}
\end{figure}
\end{center}

The wave vectors $\kappa_{\alpha}$ of equation \eqref{eq_distributions}, by virtue of equations \eqref{eq_gud_large} and \eqref{eq_guu_large} are given by
\begin{equation}\label{eq_wave_vectors}
\kappa_{\alpha} = k_F,
\end{equation}
for $\alpha= \text{BCS}, \, \uparrow \downarrow , \uparrow \uparrow$. This was determined numerically for $\kappa_{\text{BCS}}$, as shown with the (red) dots in Fig \ref{fig_phases} (a). Here we can see a difference with the 3D system, where the wave vectors are similar to $k_F$ in the BCS limit, as expected, and decrease as we move to the BEC side \cite{Obeso-Romero}, while in the 2D system they remain equal to $k_F$ throughout the crossover, which is the same as the free gas behavior, see equation \eqref{eq_f_corr_parallel_free}. As mentioned, this is expected only in the BCS limit. Also, the constant behavior of the wave vectos $\kappa_{\alpha}$ indicate that the convergence of $|\phi_{\text{BCS}}(r)|^2$ and $G_{\uparrow \downarrow}(r)$ towards a bound state distribution on the BEC side is slower in 2D than in 3D. Without formality, this can be regarded in Fig. \ref{fig_vksuks} (d), where $v_k u_k$ is still different from $v_k / u_k$ in 2D, while in 3D the difference is negligible, both plots correspond to the same numerical value of $\epsilon_{\text{spec}} / \epsilon_F$ and $ \epsilon_{\text{spec}}^{3\text{D}} / \epsilon_F^{3 \text{D}} $.\\

Regarding the phases, it is appropiate to study the large-distance phase differences between distributions. With equations \eqref{eq_gud_large} and \eqref{eq_guu_large} it is readily proven that the phase difference between the correlation functions is constant throughout the crossover:
\begin{equation}
\varphi_{\uparrow \uparrow} - \varphi_{\uparrow \downarrow} = \frac{\pi}{2} + m \pi,
\end{equation}
where $m$ is an integer. The phase differences between the pair wave function and the correlation functions are shown in Fig. \ref{fig_phases}. As expected from equations \eqref{eq_asymp_bcs_wave-function} and \eqref{eq_gud_large}, on the BCS side the nodes of the pair wave function $|\phi_{\text{BCS}}(r)|^2$ are near the nodes of $G_{\uparrow \downarrow}(r)$. Moving towards the BEC limit, after the chemical potential becomes negative, the nodes of the pair wave function approach the nodes of $G_{\uparrow \uparrow}(r)$. On the deep BEC limit the nodes of the pair wave function should approach the ones of $G_{\uparrow \downarrow}(r)$ again.

\subsection{Mean pair radius and correlation lengths}\label{sec_radius-correlation-lengths}

The exponential decay lengths are a property of the large-distance behavior of the distributions. However, to characterize the global properties of the two-body distributions, it is adequate to consider a length given by their second moment:
\begin{equation}\label{eq_rms}
\xi_{\alpha}^2 = \frac{|\int r^2 \rho_{\alpha}(r) d^2 r|}{ |\int  \rho_{\alpha}(r) d^2 r| },
\end{equation}
where $\rho_{\alpha} (r) = |\phi_{\text{BCS}}(r)|^2, G_{\uparrow \downarrow}(r) , G_{\uparrow \uparrow}(r)$, see equations \eqref{eq_pair_w_function}, \eqref{eq_antiparallel-corr-int}, and \eqref{eq_parallel-corr-int}. For the pair wave function $\phi_{\text{BCS}}(r)$ we call $\xi_{\text{BCS}}$ the mean pair radius, while the lengths $\xi_{\uparrow \downarrow}$ and $\xi_{\uparrow \uparrow}$ are called correlation lengths. The correlation length $\xi_{\uparrow \downarrow}$ has been studied in several references \cite{Randeria,Marini,Casas,Marsiglio,Yerin}, here we include it for completeness. These lengths are calculated easier using the wave vector representation, as has been done for $\xi_{\uparrow \downarrow}$ \cite{Randeria,Marini,Casas}. By means of elementary integration techniques we get
\begin{equation}\label{eq_corr_length_pair}
\xi_{\text{BCS}}^2 = \frac{\hbar^2}{m \Delta} \frac{\displaystyle  [ -1+2x^2 +2 x \sqrt{1+x^2} - x \pi +2 \text{arcsinh}(x) - 2x \, \text{arctan}(x) + \text{ln}(4+4x^2)   ]}{x + \frac{2}{3}x^3 + \frac{2}{3} (1 + x^2)^{3/2} },
\end{equation}

\begin{equation}\label{eq_corr_length_anti}
\xi_{\uparrow \downarrow}^2 = \frac{\hbar^2}{4 m \Delta} \Bigg[  x + \frac{2 + x^2}{(1+x^2)} \Bigg(   \frac{\pi}{2} +  \text{arctan}(x)  \Bigg)^{-1}   \Bigg],
\end{equation}

\begin{equation}\label{eq_corr_length_parallel}
 \xi_{\uparrow \uparrow}^2 = \frac{\hbar^2}{8 m \Delta} \frac{ 4+3 x [ \pi + x(2+\pi x ) ] + 6(x + x^3) \text{arctan}(x) }{\displaystyle (1+x^2) \bigg(  \frac{\pi}{2} + \text{arctan}(x) \bigg)  },
\end{equation}
where $x = \mu / \Delta$. These lengths are shown in Fig. \ref{fig_lengths} where, for comparison, we also plot the large-distance correlation length $\chi_{\text{spec}}= a_{2\text{D}}$. It is of interest to explore their asymptotic behavior. The asymptotic behaviors of $\xi_{\uparrow \downarrow}$ have been reported in \cite{Randeria} and \cite{Casas}. On the BCS side, $\text{ln}(1/k_Fa_{2\text{D}}) \rightarrow - \infty$, we have
\begin{equation}
\begin{split}
\xi_{\text{BCS}} &\approx \sqrt{6} \frac{1}{k_F},\\
\xi_{\uparrow \downarrow}  &\approx \frac{1}{2 \sqrt{2}} a_{2 \text{D}},\\
\xi_{\uparrow \uparrow}  &\approx \frac{1}{2} \sqrt{\frac{3}{2}} a_{2 \text{D}}.
\end{split}
\end{equation}
As seen in Fig \ref{fig_lengths}, the correlation lengths $\xi_{\uparrow \downarrow}$ and $\xi_{\uparrow \uparrow}$ increase in the BCS limit, but are smaller than $\chi_{\text{spec}}$, due to spatial oscillations. The mean pair radius $\xi_{\text{BCS}}$ tends to a finite value, that depends only on the density through $\epsilon_F$, showing a similar behavior as in 3D \cite{Ortiz}. Its value agrees well with the asymptotic behavior shown in equation \eqref{eq_asymp_bcs_wave-function}, in accordance with an apparent recovery of scale invariance. In contrast with 3D, the 2D scattering length increases on the BCS side, $a_{2\text{D}} \rightarrow \infty$, keeping a dominant role on the determination of density fluctuation sizes, while in 3D the scattering length $a_{3\text{D}} \rightarrow 0^{-}$. On the BEC side the asymptotic behaviors are
\begin{equation}
\begin{split}
\xi_{\text{BCS}} &\approx \sqrt{\frac{2}{3}} a_{2\text{D}},\\
\xi_{\uparrow \downarrow}  &\approx \sqrt{\frac{2}{3}} a_{2\text{D}},\\
\xi_{\uparrow \uparrow}  &\approx \sqrt{\frac{4}{5}} (k_F a_{2 \text{D}}) a_{2\text{D}}.
\end{split}
\end{equation}
This dependence can be seen in Fig. \ref{fig_lengths} where the departure of $\xi_{\uparrow \uparrow}$ from $\chi_{\text{spec}}$ is evident. In the 3D system it was found that in the BEC limit $\xi_{\text{BCS}}^{3\text{D}}$, $\xi_{\uparrow \downarrow}^{3\text{D}}$ and $\sqrt{2} \chi_{\text{spec}}^{3 \text{D}}$ have the same asymptotic behavior, showing a direct relation with the binding energy of a diatomic molecule \cite{Obeso-Romero,Ortiz,Palestini-Strinati-length}. In contrast, in 2D, the lengths $\xi_{\text{BCS}}$ and $\xi_{\uparrow \downarrow}$ differ from $\chi_{\text{spec}}$ by the same numerical factor. As in 3D, the correlation length $\xi_{\uparrow \uparrow}$ vanishes faster than the other lengths in the BEC limit. This means Pauli-blocking correlations become negligible, owing to the formation of molecules \cite{Obeso-Romero}. Also, as expected, on the BEC side the scattering length $a_{2\text{D}}$ determines the size of density fluctuations, which is similar to the 3D case.

\section{Final remarks}\label{sec_conclusions}

We have addressed the problem of studying the pairing mechanism in a two-dimensional homogeneous balanced mixture of two fermionic species in the BEC-BCS crossover at the mean-field level with zero temperature. This was done by analyzing the density-density correlation functions and the variational pair wave function. The analysis of the density-density correlation functions was performed by means of explicit expressions, offering a clear view of the spatial properties of the BCS-Leggett approach in 2D. These properties might give physical insight for recent variational approaches achieved with Monte-Carlo simulations \cite{Galea-Monte_Carlo,Lukas-Monte_Carlo}. Particularly, a large-distance exponential decay might be included to model the binding properties measured in experiments \cite{schunckdetermination}. To the best of our knowledge, the spatial oscillations in the density-density correlation functions have not been observed \cite{Murthy_bkt}. However, with recent advances in the experimental resolution \cite{holten2021observation}, it might be possible to find them as a signature of the quantum regime or discard them as a property that is impossible to observe.  We believe an explanation of the constant behavior of the spatial oscillation frequency in 2D requires further considerations, such as the use of a finite range interaction, instead of the contact interaction \cite{Parish-Mihaila,Neri}, and the inclusion of beyond mean-field corrections. The phase differences between the three functions show that their nodes (or maximums also) form a structure of concentric circles. In contrast with 3D, the positions of the nodes of the correlation functions never change throughout the crossover. However, the nodes of the pair wave function move from the zeros of $J_2(k_Fr)$, in the BCS limit, to the zeros of $J_0(k_F r)$, in the BEC limit.\\

Recently, the algebraic decay associated with the BKT mechanism that allows the existence of superfluidity in 2D has been observed in the density-density correlation function of parallel spins for atoms in an inhomogeneous trap, as reported in Ref. \cite{Murthy_bkt}. Our results can not be compared with those measurements due to mean-field limitations and the homogeneity considered here. At strictly zero temperature the system can exhibit off-diagonal long-range order with a finite condensed fraction. To describe the presence of a superfluid it is necessary to introduce quantum fluctuations (beyond mean-field corrections) \cite{bkt_Bighin-Salasnich,Taylor-pairing-fluctuations}. For example, within a Gaussian approximation for the fluctuations around the mean-field order parameter, a Nambu-Goldstone field $\theta(\mathbf{r})$ has to be considered with the order parameter (the Bose pairing field) in order to describe superfluidity via the BKT mechanism \cite{bkt_Bighin-Salasnich,Taylor-pairing-fluctuations,Mulkerin_bkt}. This introduces a phase-phase correlator, which exhibits the algebraic decay associated to quasi-long range order \cite{bkt_Bighin-Salasnich}. In our approach the phase is constant $\theta (\mathbf{r})=0$, making the phase-phase correlator independent of the position, as expected \cite{bkt_Bighin-Salasnich}.
In general, quantum fluctuations become important in 2D, specially for non-zero temperatures. In the BCS limit, at zero temperature, we expect corrections to modify slightly the large-distance behavior \cite{Salasnich-gaussian}. Instead, in the BEC limit quantum fluctuations allow to obtain the correct behavior of an interacting gas of bosonic molecules \cite{Salasnich-gaussian}. Therefore, the large-distance correlation length might still be related to the binding properties of a molecule, but we expect the frequency of spatial oscillations to decrease in the BEC limit, in accordance with the formation of a two-body bound state. It will be of interest to compare the behavior of phase-phase correlators with the density-density correlation functions at finite temperature to acquire a wide view of the conditions that allow the existence of superfluidity. For example, in the deep BCS limit, we can identify two temperature limits where the system loses a scale: The critical temperature $T_{\text{BKT}}$ and zero temperature. Then, it might be interesting to explore experimentally how the system evolves between these two limits, although the BCS limit is hard to obtain experimentally \cite{schunckdetermination,Bloch-review}.\\

Let us remark some similarities and differences between 2D and 3D, at the mean-field level. Firstly, from dimensional analysis, the interaction strength $g$ in 2D is unable to provide a natural length associated to the contact interaction, differently from 3D. Still, the anomalous breaking of scale invariance, a feature of the 2D contact interaction, allows to introduce the $s$-wave scattering length. In 2D and 3D, the scattering lengths dominate the many-body properties due to the renormalization procedure, which, in 2D, coincidentally allows to solve the gap and number equations. Other important aspect is the presence of a bound state energy for any interaction strength throughout the crossover in 2D, while in 3D a bound state appears only for positive scattering length (BEC side). Related to this behavior, it is found that in 2D the binding energy per pair $\epsilon_b$ is equal to the threshold energy required to create a quasiparticle with minimum momentum transfer $\epsilon_{\text{spec}}$, while in 3D they are different, $ \epsilon_b^{3\text{D}} \neq  \epsilon_{\text{spec}}^{3 \text{D}} $, except in the BEC limit. However, a similarity that has been demonstrated here is that the large-distance exponential decay of the density-density correlation functions and the pair wave function in 2D and 3D is determined by $\epsilon_{\text{spec}}$ and $\epsilon_{\text{spec}}^{3 \text{D}}$ respectively, from where we identified the large-distance correlation lengths, $\chi_{\text{spec}}$ and $\chi_{\text{spec}}^{3\text{D}}$ , given in terms of the gap and chemical potential, see eqs. \eqref{eq_chi_spec_2d} and \eqref{eq_chi_3d}. In 2D the large-distance correlation length is equal to the scattering length $\chi_{\text{spec}}= a_{2\text{D}}$, while in 3D $\chi_{\text{spec}}^{3\text{D}} \approx a_{3 \text{D}}$ in the BEC limit only. A notorious difference in the large-distance behavior of the correlation functions in 2D and 3D lies in the behavior of the frequencies of spatial oscillations. In 2D it is constant throughout the crossover, being equal to the Fermi wave number, equation \eqref{eq_wave_vectors}. In 3D the respective frequencies are similar to the Fermi wave number in the BCS limit and decrease as we approach the BEC limit. A striking similarity is shown in Fig. \ref{fig_vksuks} when analyzing the behavior of the variational parameters as functions of the norm of the wave vector, it is difficult to see the differences. Differently, in position space their respective Fourier transforms have characteristic features depending on the dimension. Regarding the second moments of the distributions, also called correlation lengths and mean pair radius, see eq. \eqref{eq_rms}, we have shown that the information about the large-distance structure is quite diluted in these lengths. We found that they behave similarly to their analogues in 3D \cite{Palestini-Strinati-length,Randeria,Ortiz}. In the BCS limit the correlation lengths diverge, while the mean pair radius is finite. Instead, in the BEC limit the three lengths tend to zero. However, a difference with 3D is that in 2D the correlation length of unlike species $\xi_{\uparrow \downarrow}$ and the mean pair radius $\xi_{\text{BCS}}$ are not strictly equal to the large-distance correlation length $\chi_{\text{spec}}$ in the BEC limit.\\

As an extension of this work, it will be of interest to study the large-distance behavior of the correlation functions in quasi-2D geometries, which are of current interest for understanding the evolution from 3D to 2D \cite{Levinsen-Parish,Dyke_E,Toniolo,Ries_quasi2d}. Also, it is of interest to see how the large-distance properties are modified by beyond mean-field corrections \cite{Bighin,Bighin-Salasnich-ftqf,Marsiglio,He}, the use of a short-range potential \cite{Parish-Mihaila,Caballero,Neri}, and with different pairing mechanisms \cite{Yerin,Camacho,Dominguez-Castro}. Particularly, the study of correlation lengths is important to appreciate the BEC-BCS crossover in solid-state systems \cite{Nakagawa,Suzuki}. An important aspect towards achieving the BCS limit concerns the increase of the large-distance correlation lengths, which can become of macroscopic size, indicating that experiments require huge systems. Also, they indicate that care must be taken when using local-density approximations in the BCS limit \cite{Obeso-Romero}.\\

\begin{acknowledgments}

We thank support from UNAM PAPIIT-IN108620. J.C.O.-J. acknowledges support from a CONACYT scholarship.

\end{acknowledgments}

\appendix

\section{Inverse Fourier transforms of $g_{\uparrow \downarrow}(r)$ and $g_{\uparrow \uparrow}(r)$}\label{ap_inv_fourier_transf}
The inverse Fourier transform of $g_{\uparrow \downarrow}(r)$ can be written in the following way:
\begin{equation}\label{eq_ap_fourier_ud}
\int d^2 r \; e^{- i \mathbf{k} \cdot \mathbf{r}} g_{\uparrow \downarrow}(r) = \frac{(2 \pi)}{k_F^2} \int_{0}^{\infty} g_{\uparrow \downarrow} (\rho) \rho J_0 (\kappa \rho) d \rho,
\end{equation}
where we made a transformation to polar coordinates and evaluated the angular integral, which allows us to identify the integral representation of the Bessel function of the first kind of order zero $J_0(\kappa \rho)$ \eqref{eq_besselj0}. Also, we scaled variables with the Fermi wave number $k_F$, such that $\rho = k_F r$ and $\kappa = k / k_{F}$, where $k = |\mathbf{k}|$, while energies will be scaled with the Fermi energy, $\tilde{\Delta} = \Delta / \epsilon_F$, $\tilde{\mu} = \mu / \epsilon_F$.
Substituting the explicit form of $g_{\uparrow \downarrow}(r)$ given in equation \eqref{eq_gud} we get
\begin{equation}\label{eq_ap_fourier_ud2}
\int d^2 r \; e^{- i \mathbf{k} \cdot \mathbf{r} } g_{\uparrow \downarrow}(r) = \frac{\tilde{\Delta}}{2} \int_{0}^{\infty} \rho J_0(\rho) K_{0} \Bigg( \frac{\rho}{ k_F a_{2\text{D}}}  \Bigg) J_{0}(\kappa \rho) d\rho.
\end{equation}
The integral on the right side of equation \eqref{eq_ap_fourier_ud2} has been evaluated and can be found in a table of integrals, like \cite{Gradshteyn} or \cite{Bateman-tables}. Here we write the general expression:
\begin{equation}
\int_0^{\infty} x J_0(ax) K_0(bx) J_0(cx) \; dx = [a^4 + b^4 +c^4 -2a^2c^2 + 2a^2b^2 + 2b^2c^2]^{-1/2}
\end{equation}
where $\text{Re }b > |\text{Im } a|$ and $c>0$. With the identification of $a=1$, $b= \{ [(\tilde{\mu}^2 + \tilde{\Delta}^2 )^{1/2} - \tilde{\mu}]/2 \}^{1/2}$, and $c=\kappa$, together with the aid of equations \eqref{eq_chem_ea} and \eqref{eq_gap_ea}, we get the desired result:
\begin{equation}
\int d^2 r \; e^{- i \mathbf{k} \cdot \mathbf{r}} g_{\uparrow \downarrow}(r) = \frac{\tilde{\Delta}}{2 \sqrt{ (\kappa^2 - \tilde{\mu})^2 + \tilde{\Delta}^2  }} = v_k u_k.
\end{equation}

In a similar way, we find the inverse Fourier transform of $g_{\uparrow \uparrow}(r)$ to be given by
\begin{equation}\label{eq_inv_fourier_guu}
\int d^2 r \; e^{- i \mathbf{k} \cdot \mathbf{r} } g_{\uparrow \uparrow}(r) = \frac{\tilde{\Delta}}{2} \int_0 ^{\infty} K_{1}\Bigg( \frac{\rho}{ k_F a_{2\text{D}}} \Bigg) \, J_{1}(\rho) \rho J_0(\kappa \rho) d \rho.
\end{equation}
The integral on the right side can be identified as a particular case of a great variety of integrals, see for instance \cite{Gradshteyn} and \cite{Bateman-tables}. However, not all the expressions are adequate for the purpose of identifying $v_k^2$. Thus, we will give a sketch of how to prove that the right side of equation \eqref{eq_inv_fourier_guu} is $v_k^2$. We can start with the following general integral \cite{Gradshteyn,Bateman-tables}:
\begin{equation}\label{eq_transf_general_int}
\int_0^{\infty} x^{\nu + 1} K_{\mu}(ax) \, I_{\mu}(bx) \, J_{\nu}(cx) dx = \frac{(ab)^{-\nu-1} c^{\nu} e^{-(\nu + 1/2)\pi i}  Q_{\mu-1/2}^{\nu+1/2}(u)}{\sqrt{2 \pi} (u^2 - 1)^{ \frac{1}{2}\nu + \frac{1}{4} }}
\end{equation}
where $u = (a^2 + b^2 + c^2)/(2ab)$, with $\text{Re }a  > |\text{Re }b | + |\text{Im }c |$, $\text{Re }\nu > -1$ and $\text{Re}(\mu + \nu) > -1$. In this equation $I_{\mu}$ is the modified Bessel function of the first kind of order $\mu$ and $Q^{\beta}_{\alpha}$ is an associated Legendre function of the second kind. The values of the constants we are interested in are $a =  \{ [(\tilde{\mu}^2 + \tilde{\Delta}^2 )^{1/2} - \tilde{\mu}]/2 \}^{1/2} $, $b=i$ and $c=\kappa$, while the subscripts are $\mu = 1$ and $\nu = 0$. The associated Legendre function of the second kind can be expressed in terms of the hypergeometric function $_2F_{ 1}$ in the following way \cite{Bateman,Gradshteyn}:
\begin{equation}\label{eq_assoc_legendre_2}
Q_{\alpha}^{\beta} (z) = \frac{e^{\beta \pi i}}{2^{\alpha +1}} \frac{\Gamma (\alpha + \beta + 1)}{\Gamma (\alpha + 3/2)} \frac{\Gamma (1/2) (z^2-1)^{\beta /2}}{z^{\alpha + \beta +1}} \hspace{0.5cm} F_{\hspace{-0.5cm}2 \hspace{0.4cm} 1} \Bigg( \frac{\alpha + \beta }{2} + 1, \frac{\alpha + \beta + 1}{2} ; \alpha + \frac{3}{2}, \frac{1}{z^2} \Bigg),
\end{equation}
where $\Gamma$ is the well-known Gamma function. Finally, we also need an integral representation of the hypergeometric function \cite{Bateman,Gradshteyn}:
\begin{equation}\label{eq_rep_integral_2f1}
F_{\hspace{-0.5cm}2 \hspace{0.4cm} 1} (\alpha, \beta, \gamma; z) = \frac{1}{B(\beta, \gamma - \beta)} \int_0^{1} t^{\beta -1}(1-t)^{\gamma - \beta -1}(1-tz)^{- \alpha} \; dt,
\end{equation}
where $\text{Re } \gamma > \text{Re }\beta > 0 $, and $B$ is the beta function. For our particular case, the integral in equation \eqref{eq_rep_integral_2f1} can be evaluated by elementary integration techniques. Hence, we can substitute equations \eqref{eq_assoc_legendre_2} and \eqref{eq_rep_integral_2f1} into \eqref{eq_transf_general_int} to obtain
\begin{equation}\label{eq_paso_trans_hankel_paralelos}
\begin{split}
\int_0^{\infty} x \; K_{1}\Bigg( \frac{x}{\sqrt{2} k_F \chi_b} \Bigg) \, I_{1}(ix) \, J_{0}(\kappa x) \; dx =  \frac{1}{i 2 \sqrt{ (1- \tilde{\mu})}} \;      \Bigg( \frac{1}{ \Big( 1- \frac{1}{u^2}  \Big)^{1/2}} - 1  \Bigg),
\end{split}
\end{equation}
where $u = (\kappa^2 - \tilde{\mu})/(2i \sqrt{1- \tilde{\mu}})$ and we expressed the gap $\tilde{\Delta}$ in terms of the chemical potential $\tilde{\mu}$ using equations \eqref{eq_chem_ea} and \eqref{eq_gap_ea}. From equation \eqref{eq_paso_trans_hankel_paralelos} we can identify the right side of equation \eqref{eq_inv_fourier_guu} recalling that $I_{\nu}(x) = e^{- \nu \pi i /2} J_{\nu}(e^{i \pi / 2} x)$. After some rearrangements we conclude that
\begin{equation}
\int d^2 r \; e^{- i \mathbf{k} \cdot \mathbf{r} } g_{\uparrow \uparrow}(r) = v_k^2.
\end{equation}

\section{Large-distance approximation of the pair wave function}\label{ap_large-distance-pair}

\begin{center}
\begin{figure}[h!]
\includegraphics[width=0.5\linewidth]{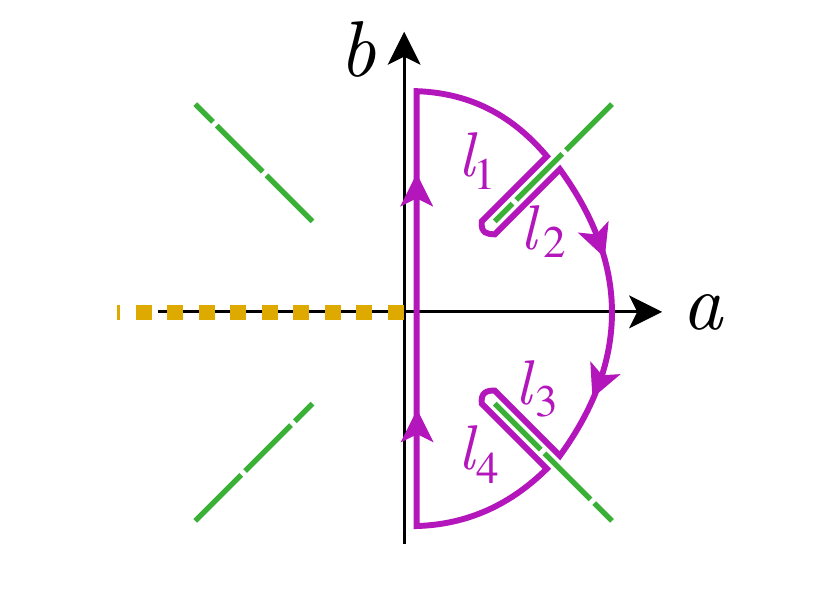}
\caption{(Color online) Illustration of the branch cuts of the integrand in equation \eqref{eq_ap_int_pair_wave}. The large dashes (green) correspond to the branch cuts of $\mathcal{F}(ix)$, see equation \eqref{eq_ap_pair_wave}, while the short dashes (orange) to the branch cut of $\sqrt{x}$. The solid line (purple) is the contour used in the Cauchy's integral formula in equation \eqref{eq_cauchy_formula}. Close to the branch cuts we have four paths denoted by $l_i$, with $i=1,2,3,4$.} \label{fig_branch_cuts}
\end{figure}
\end{center}

Introducing the approximation of the Bessel function $J_0(kr)$, given in equation \eqref{eq_aprox_Bessel}, into equation \eqref{eq_hankel_trans} we have
\begin{equation}
\phi_{\text{BCS}}(r) \approx \frac{k_{\Delta}^{3/2}}{\sqrt{2 \pi^3 r}} \Bigg[  \frac{S_{+}(r) e^{-i\pi /4} + S_{-}(r)e^{i \pi /4}}{2}   \Bigg],
\end{equation}
where we scaled the lengths with the wave vector $k_{\Delta}$ associated to the gap $\Delta = \hbar^2 k_{\Delta}^2/2m $ and defined
\begin{equation}
S_{\pm}(r) = \int_0^{\infty} \sqrt{p} \mathcal{F}(p) e^{\pm i p k_{\Delta} r} dp.
\end{equation}
For the pair wave function we have
\begin{equation}\label{eq_ap_pair_wave}
\mathcal{F}(p) = \sqrt{(p^2 - \mu_\Delta)^2 + 1} - (p^2 - \mu_\Delta),
\end{equation}
where $\mu_\Delta = \mu / \Delta$. Using the change of variable $p = e^{i \pi /2} x$ in $S_{+}(r)$ and $p = e^{-i \pi/2}x$ in $S_{-}(r)$ we get
\begin{equation}\label{eq_ap_int_pair_wave}
\phi_{\text{BCS}}(r) \approx - \frac{i k_{\Delta}^{3/2}}{2^{3/2}\sqrt{ \pi^3 r}} \Bigg[  \int_{\mathbb{I}} \sqrt{x} \mathcal{F}(ix) e^{- x k_{\Delta} r} dx  \Bigg],
\end{equation}
where the integral is over the imaginary axis, from $x=- i\infty$ to $x=i \infty$. The integrand has five branch cuts, see equations \eqref{eq_ap_pair_wave} and \eqref{eq_ap_int_pair_wave}. Four of them are determined by the equations:
\begin{equation}
\text{Re}[(x^2 + \mu_{\Delta})^2 + 1] \leq 0, \text{ and Im}[(x^2 + \mu_{\Delta})^2 + 1]=0.
\end{equation}
Using $x = a + ib$ we find that those branch cuts correspond to points in the hyperbola $b^2 - a^2 = \mu_{\Delta}$ whose magnitude satisfies $|x|^2 \geq (\mu_{\Delta}^2 + 1)^{1/2}$. As the integrand in the right side of equation \eqref{eq_ap_int_pair_wave} decreases exponentially when $\text{Re }x \rightarrow \infty$, we can close a contour to the right side of the complex plane with a semicircle-like contour, which surrounds infinitesimally two branch cuts. We illustrate this contour in Fig. \ref{fig_branch_cuts}

From Cauchy's integral formula we have \cite{Marsden}
\begin{equation}\label{eq_cauchy_formula}
\oint_{\mathcal{C}} \sqrt{x} \mathcal{F}(ix) e^{- x k_{\Delta} r} dx  =0,
\end{equation}
where $\mathcal{C}$ is the contour shown in Fig. \ref{fig_branch_cuts}. Taking the radius of the semicircle to infinity, we get
\begin{equation}\label{eq_ap_int_parametric1}
\begin{split}
\int_{\mathbb{I}} \sqrt{x} \mathcal{F}(ix) e^{- x k_{\Delta} r} dx  = & 2 \int_{l_2} \sqrt{x} \sqrt{|(x^2 + \mu_{\Delta})^2 +1  |}e^{-i\pi /2}  e^{-x k_{\Delta} r} dx\\
 &+ 2 \int_{l_4} \sqrt{x} \sqrt{|(x^2 + \mu_{\Delta})^2 +1  |}e^{-i\pi /2}  e^{-x k_{\Delta} r} dx,
\end{split}
\end{equation}
where $l_2$ and $l_4$ are the trajectories depicted in Fig. \ref{fig_branch_cuts}. The parametrization of $l_2$ is given by $\gamma_2(t) = t+ i (t^2 + \mu_{\Delta})^{1/2}$, while the parametrization of $l_4$ is $\gamma_4(t) = t-i (t^2 + \mu_{\Delta})^{1/2} $,  with $t \in [t_0, \infty )$, where
\begin{equation}
t_0 = \Bigg( \frac{(\mu_{\Delta}^2 + 1)^{1/2} - \mu_{\Delta}}{2}  \Bigg)^{1/2}.
\end{equation}
With the explicit form of the parametrizations we can join the two integrals in equation \eqref{eq_ap_int_parametric1} in the following way:
\begin{equation}\label{eq_ap_int_parametric2}
\int_{\mathbb{I}} \sqrt{x} \mathcal{F}(ix) e^{- x k_{\Delta} r} dx  = 4 \int_{t_0}^{\infty}  \sqrt{4t^2(t^2 + \mu_{\Delta})-1}e^{-i\pi /2}  e^{-t k_{\Delta} r} \text{Re} [e^{-i \sqrt{t^2 + \mu_{\Delta}}} \sqrt{\gamma_2(t)} \gamma_2'(t)] dt.
\end{equation}
To get an explicit expression we can notice that the branch cuts always remain in their own quadrant. Then with de Moivre's formula the parametrization can be written as
\begin{equation}
\gamma_2(t) = (2t^2 + \mu_{\Delta})^{1/2} [\text{cos } \theta(t) + i \text{sin } \theta(t) ],
\end{equation}
where we have defined
\begin{equation}
\theta (t) = \text{arctan} \Bigg( \frac{\sqrt{t^2 + \mu_{\Delta}}}{t} \Bigg).
\end{equation}
This form helps us to calculate $\sqrt{\gamma_2(t)}$ in equation \eqref{eq_ap_int_parametric2}. Then we get explicitly
\begin{equation}\label{eq_ap_final_int}
\begin{split}
\int_{\mathbb{I}} \sqrt{x} \mathcal{F}(ix) e^{- x k_{\Delta} r} dx  =& 4 \int_{t_0}^{\infty}  \sqrt{4t^2(t^2 + \mu_{\Delta})-1}\; e^{-i\pi /2}  e^{-t k_{\Delta} r}  (2t^2 + \mu_{\Delta})^{1/4}\\
&\times \Bigg\{ \text{cos}(\sqrt{t^2 + \mu_{\Delta}})  \Bigg[  \text{cos}(\theta (t)/2) - \frac{t \text{sin}(\theta (t)/2)}{\sqrt{t^2 + \mu_{\Delta}}} \Bigg]\\
& +  \text{sin}(\sqrt{t^2 + \mu_{\Delta}}) \Bigg[  \frac{t \text{cos} (\theta (t)/2)}{\sqrt{t^2 + \mu_{\Delta}}} + \text{sin}(\theta (t)/2) \Bigg]  \Bigg\}dt .
\end{split}
\end{equation}
Substituting equation \eqref{eq_ap_final_int} into equation \eqref{eq_ap_int_pair_wave} and scaling variables with $k_F$ instead of $k_{\Delta}$ we obtain
\begin{equation}\label{eq_asymp_pair_anti}
\begin{split}
\phi_{\text{BCS}}(r)  \propto \frac{1}{\sqrt{k_F r}} \int_{\tau_0}^{\infty} e^{- \tau k_F r} \Bigg( \frac{2 \tau^2 + \tilde{\mu}}{\tilde{\Delta}} \Bigg)^{1/4} \Bigg[ \text{cos}\Bigg( \sqrt{\frac{\tau^2 + \tilde{\mu}}{\tilde{\Delta}}} \Bigg) \Bigg( \text{cos} (\theta(\tau)/2) - \frac{\tau \text{sin}(\theta (\tau) /2)}{\sqrt{\tau^2 + \tilde{\mu}}}  \Bigg)  \\
  +   \text{sin}\Bigg( \sqrt{\frac{\tau^2 + \tilde{\mu}}{\tilde{\Delta}}} \Bigg) \Bigg( \frac{\tau \text{cos}(\theta (\tau) / 2)}{\sqrt{\tau^2 + \tilde{\mu}}}  + \text{sin}(\theta (\tau) / 2)   \Bigg)  \Bigg] \Bigg(  \frac{\sqrt{4 \tau^2 (\tau^2 + \tilde{\mu} ) - \tilde{\Delta}^2}}{\tilde{\Delta}} \Bigg) \; d \tau,
\end{split}
\end{equation}
where $\tau_0 = \{ [(\tilde{\mu}^2 + \tilde{\Delta}^2)^{1/2} - \tilde{\mu} ]/2 \}^{1/2}$, and we have introduced the function
\begin{equation}
\theta (\tau) = \text{tan}^{-1} \Bigg( \frac{\sqrt{\tau^2 + \tilde{\mu}}}{\tau} \Bigg).
\end{equation}
The approximation of the Bessel function in equation \eqref{eq_aprox_Bessel} removes part of the structure, mainly the oscillations. Nevertheless, the main aspect to point out is that the integrand has an exponential factor $\text{exp}(- \tau k_F r)$. This factor should persist after integration, but evaluated at $\tau_0$. Hence, we expect the Hankel transform of $v_{k}/u_{k}$ to have an exponential decay behavior of the form of equation \eqref{eq_fr-theo}.

\bibliography{bibliography-2d}

\end{document}